\documentclass[aip,amsmath,amssymb,reprint]{revtex4-1}

\usepackage[utf8]{inputenc}
\usepackage[T1]{fontenc}
\usepackage{mathptmx}
\usepackage{graphicx}% Include figure files
\usepackage{dcolumn}% Align table columns on decimal point
\usepackage{bm}% bold math
\usepackage{graphicx}  % needed for figures
\usepackage{dcolumn}   % needed for some tables
\usepackage{bm}        % for math
\usepackage{amssymb}   % for math
\usepackage{amsmath}
\usepackage{color}

\newcommand{\ket}[1]{\left| #1 \right\rangle}
\newcommand{\qo}[1]{``#1''}
\newcommand{\COR}[1]{\textcolor{black}{#1}}

\begin{document}
%\preprint{AIP/123-QED}

\title[Nonlocal Quantum Erasure of Phase Objects]{Nonlocal Quantum Erasure of Phase Objects}

\author{Lu \surname{Gao}}
\thanks{L. Gao and Y. Zhang contributed equally to this work.}
\affiliation{School of Science, China University of Geosciences, Beijing 100083, China}

\author{Yingwen \surname{Zhang}}
\email{Yingwen.Zhang@nrc-cnrc.gc.ca}
\affiliation{Physics Department, University of Ottawa, Advanced Research Complex, 25 Templeton Street, Ottawa ON Canada, K1N 6N5}
\affiliation{National Research Council of Canada, 100 Sussex Drive, Ottawa ON Canada, K1A0R6}

\author{Eliahu \surname{Cohen}}
\affiliation{Physics Department, University of Ottawa, Advanced Research Complex, 25 Templeton Street, Ottawa ON Canada, K1N 6N5}
\affiliation{Faculty of Engineering and the Institute of Nanotechnology and Advanced Materials, Bar Ilan University, Ramat Gan 5290002, Israel}

\author{Avshalom C. \surname{Elitzur}}
\affiliation{Iyar, The Israeli Institute for Advanced Research, POB 651, Zichron, Ya'akov, 3095303, Israel}
\affiliation{Institute for Quantum Studies, Chapman University, Orange, CA 92866, USA}

\author{Ebrahim \surname{Karimi}}
\affiliation{Physics Department, University of Ottawa, Advanced Research Complex, 25 Templeton Street, Ottawa ON Canada, K1N 6N5}
\affiliation{National Research Council of Canada, 100 Sussex Drive, Ottawa ON Canada, K1A0R6}

\date{\today}

\begin{abstract}
The Franson interference is a fourth order interference effect, which unlike the better known Hong-Ou-Mandel interference, does not require the entangled photon pairs to be present at the same space-time location for interference to occur -- it is nonlocal. Here, we use a modified Franson interferometer to experimentally demonstrate the nonlocal erasure and correction of an image of a phase-object taken through coincidence imaging. This non-local quantum erasure technique can have several potential applications such as phase corrections in quantum imaging and microscopy and also user authentication of two foreign distant parties.

\end{abstract}

\maketitle

%\section{INTRODUCTION}

The Franson interference (FI) is a fourth-order two-particle interference effect proposed by James Franson~\cite{Franson1989} that was first demonstrated in 1990~\cite{Kwiat1990,Ou1990}. Unlike the Hong-Ou-Mandel interference effect~\cite{Hong1987}, where the two particles are required to be brought to the same space-time location for interference, the FI can be observed with the two particles at vastly separate locations. In an optical Franson interferometer, a correlated photon pair is first generated, e.g. through spontaneous parametric down-conversion (SPDC), and each photon is sent to a different path. An unbalanced Mach–Zehnder interferometer (UMZI) is constructed in the path of each photon, where the path length difference in each UMZI is longer than the coherence length of the photons. This way, no interference will be observed for each individual photon. However, when the photon pairs are detected in coincidence, interference appears only if the difference in optical path lengths in the two UMZI is smaller than the coherence length of the photons. This is the result from the interference between the two amplitudes of both photons traveling short paths (SS) of the UMZIs with the amplitudes of both photons traveling long paths (LL) of the UMZIs, as these two cases are temporally indistinguishable when measured in coincidence. However, due to the probabilistic nature of the 50:50 beamsplitter used in the UMZI, the cases where one photon travels the long path and the other the short path (LS), and \emph{vice versa} (SL), are also present. These LS and SL scenarios are temporally distinguishable and do not contribute to the interference. If the detectors do not have enough timing resolution to resolve the LS and SL cases from the LL and SS, then an interference visibility of only $50\%$ can be observed. To overcome this problem, one either has to use an UMZI with the optical path length difference longer than the timing resolution of the detectors~\cite{Kwiat1993} or use polarization interferometers with polarization entangled photons~\cite{Kim2017}.

FI has been demonstrated at large distances between the two interferometers through free-space~\cite{Franson1991} and fibre~\cite{Rarity1992,Tapster1994}. This has led to the proposal and demonstration of time-bin encoded quantum key distribution (QKD) using FI~\cite{Ekert1992,Brendel1999,Marcikic2002,Halder2007,Dynes2009}. This nonlocal property of FI had also been used in testing the speed of \qo{spooky action at a distance}~\cite{Salart2008} and nonlocal pulse shaping of entangled photons~\cite{Bellini2003}. Proposals and demonstrations of partial quantum measurement reversal employing a modified Franson interferometer have also been shown~\cite{Elitzur2001,Elitzur2011,Xu2011}. Similar schemes employing superconducting qubits are known as well~\cite{Paraoanu2006,Katz2006}.

In this work, we demonstrate the nonlocal erasure of an image of a phase object taken with a coincidence imaging setup~\COR{\cite{Aspden2013,Zhang2019}} by using an embedded Franson interferometer. Here, polarization entangled photon pairs are first generated through SPDC. The signal and idler photons are then sent to separate paths, each to a polarization Sagnac interferometer (SI) constructed using a polarization beamsplitter (PBS). As the photons are either horizontally ($H$) or vertically ($V$) polarized, no interference will be observed for each individual photon while FI will still be seen when post-selecting on the correct polarization basis and registering the photons in coincidence. When a phase object is inserted in one of the SI, FI will be disturbed and the object will appear on the camera through coincidence imaging. However, FI can be restored by placing an object with the same phase profile in the other SI, thus erasing the presence of both phase objects. We believe this nonlocal quantum erasure technique can have potential applications in phase corrections as part of quantum imaging and also performing user authentication of two distant parties.

%\section{Experiment}

\begin{figure}[htbp]
\centering \includegraphics[width=0.48\textwidth]{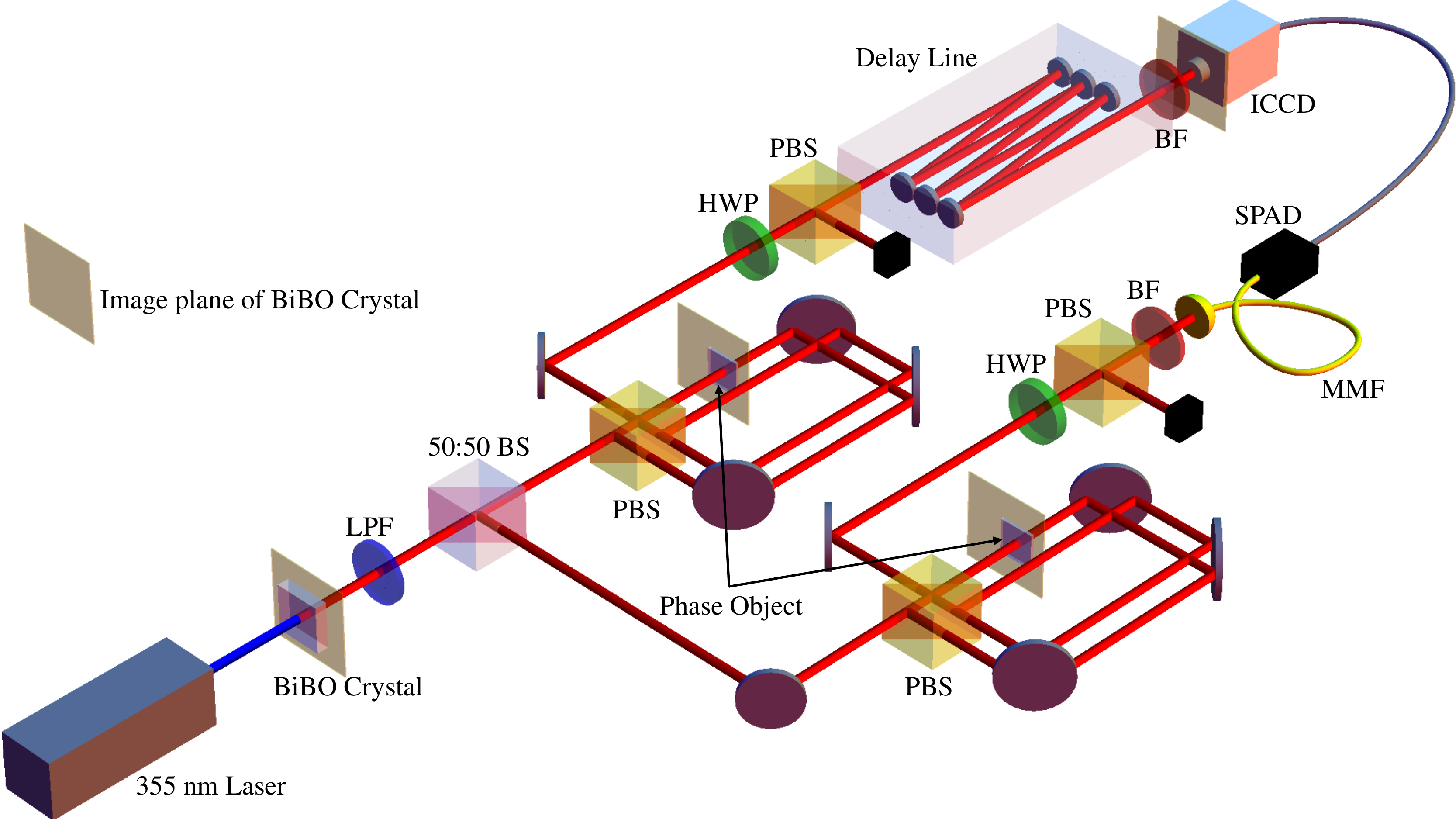}
\caption{Experimental setup for nonlocal quantum erasure of a phase object. Polarization entangled photon pairs are generated by a nonlinear crystal (BiBO). The signal and idler photons are separated and sent to two separate polarization Sagnac interferometers (SIs) where the phase object is placed. Upon exiting the SI, the photons undergo polarization post-selection at a HWP and PBS. The photons that succeed in the post-selection are transmitted by the PBS with the idler photon collected by a bucket detector and the signal photon sent through a delay line to a camera on which the image of the phase objects is to be formed. Those photons that failed the post-selection are reflected by the PBS and are ignored. Imaging lenses are not shown in the setup. Figure legends: BiBO - sandwiched bismuth triborate crystal; LPF - Long-Pass Filter; BS - Beamsplitter;	HWP - Half-wave plate; PBS - polarizing beamsplitter; BF- Bandpass Filter; MMF - Multi-Mode Fibre; SPAD - Single Photon Avalanche Diode; ICCD - Intensified CCD camera.}
\label{Fig.1}
\end{figure}

The conceptual arrangement of the nonlocal quantum image erasure setup is illustrated in Fig.~\ref{Fig.1}. A 100~mW, 355~nm, 10~ps pulse width, with 100~MHz repetition rate, quasi-cw laser (JDSU Xcyte CY-SM100) is used to pump two adjacent, 0.5~mm thick Type-I bismuth triborate (BiBO) crystals with their optical axes oriented perpendicular to each other to generate polarization entangled photon pairs in the state
\begin{equation} \label{1}
\ket{\psi}=\frac{1}{\sqrt{2}}\left(\ket{H}_s\ket{H}_i+e^{i\varphi}\ket{V}_s\ket{V}_i\right),
\end{equation}
with a wavelength of 710~nm~\cite{Kwiat1999}. Here, the subscripts $s$ and $i$ denote the signal and idler photon. The phase $\varphi$ is dictated by the orientation, i.e. phase-matching of the BiBO crystal. The 355~nm pump beam is afterwards filtered out with a long-pass filter and the photon pairs separated by a 50:50 \COR{beamsplitter (BS)} into two separate paths in which each contains a polarization SI constructed using a PBS. Using a polarization SI ensures overall stability and also guarantees that no second-order interference will be observed for each individual photon as the PBS will send the photons on either the clockwise or anti-clockwise path of the SIs and never as a superposition. The clockwise and anti-clockwise beam paths of the SIs are slightly displaced from each other such that the phase objects can be placed in just one of the paths. Upon exit of the SIs we obtain the state
\begin{equation} \label{2}
\ket{\psi}=\frac{1}{\sqrt{2}}\left(\ket{H}_s\ket{H}_i+e^{i\phi}\ket{V}_s\ket{V}_i\right),
\end{equation}
where $\phi=\frac{\omega_p}{2c}(\Delta L_s+\Delta L_i)+\varphi$ with $\omega_p$ being the pump frequency, $c$ the speed of light in vacuum and $\Delta L_s\,(\Delta L_i)$ is the difference in optical path length between the clockwise and anti-clockwise paths of the SI for the signal (idler) photons. In order to observe FI, $\Delta L_s-\Delta L_i$ must be smaller than the coherence length of the SPDC photons~\cite{Ou1990,Kwiat1993}. This condition is easily satisfied by using a SI.

The photons then undergo post-selection of the polarization through the combination of a half-wave plate (HWP) and a PBS, then post-selection of the wavelength through bandpass filters centered on $710\pm5$~nm. The signal photons are sent to an intensified charge-coupled device (ICCD) camera and the idler photons are collected by a bucket detector composed of a 200~$\mu$m core multi-mode fibre attached to a single photon avalanche photodiode (SPAD). The shutter of the ICCD camera is triggered by the SPAD upon detection of the idler photon in order to collect the signal photon in coincidence. A 24~m optical delay line has to be constructed in the path of the signal photon to compensate for the electronic delay inside the ICCD and SPAD~\cite{Aspden2013}. Imaging lenses (not shown in Fig.~\ref{Fig.1}) are used to ensure that the ICCD camera and the phase objects are in the image-plane of the BiBO crystal.

When setting both HWP to $22.5^\circ$ during polarization post-selection, our two photon state from Eq.~(\ref{2}) becomes
\begin{eqnarray}
\ket{\psi}=&\frac{1}{2\sqrt{2}}\left[(1+e^{i\phi})\left(\ket{H}_{s}\ket{H}_{i}+\ket{V}_{s}\ket{V}_{i}\right)\right.\nonumber\\
&+\left.(1-e^{i\phi})\left(\ket{H}_{s}\ket{V}_{i}+\ket{V}_{s}\ket{H}_{i}\right)\right].
\end{eqnarray}
As we see, when $\phi$ is varied from 0 to $2\pi$, $\ket{\psi}$ would oscillate between the state $\frac{1}{\sqrt{2}}(\ket{H}_{s}\ket{H}_{i}+\ket{V}_{s}\ket{V}_{i})$ and
$\frac{1}{\sqrt{2}}(\ket{H}_{s}\ket{V}_{i}+\ket{V}_{s}\ket{H}_{i})$.

For this experiment we initially set $\phi=2n\pi$ for $n \in \mathbb{Z}$ such that we see constructive interference when post-selecting on the $\frac{1}{\sqrt{2}}(\ket{H}_{s}\ket{H}_{i}+\ket{V}_{s}\ket{V}_{i})$ basis (by rotating both HWPs to $22.5^\circ$) and see destructive interference when post-selecting on $\frac{1}{\sqrt{2}}(\ket{H}_{s}\ket{V}_{i}+\ket{V}_{s}\ket{H}_{i})$ (by rotating one HWP to $22.5^\circ$ and the other to $-22.5^\circ$). \COR{An interesting point to note here is that $\phi$ depends on both the optical path length difference $\Delta L_s-\Delta L_i$ and the phase $\varphi$ introduced in the BiBO crystal. $\phi$ can therefore be tuned by either slightly adjusting the orientation of a mirror inside one of the SI or by adjusting the phase-matching of the BiBO crystal through the crystal orientation. This shows that the polarization FI is already capable of correcting the phase introduced by a birefringent material placed before the signal and idler photons are separated.}

When a phase object (piece of glass) with a thickness much longer than the coherence length of the SPDC photons is inserted in, say the clockwise arm of the idler photon SI, $\Delta L_s-\Delta L_i$ is no longer small and FI is lost. An image of the phase object will thus appear when we post-select on destructive interference and the shadow of the phase object will appear when post-selecting on constructive interference. However, when the same phase object is inserted in the clockwise arm of the SI for the signal photon, $\Delta L_s-\Delta L_i$ will be restored to its original value and FI will be reinstated. A slight adjustment to the insertion angle of the object will restore $\phi=2n\pi$ and the presence of the two phase objects is thus \qo{erased} nonlocally.

%\section{Results}
%
\begin{figure}[htbp]
	\centering \includegraphics[width=0.48\textwidth]{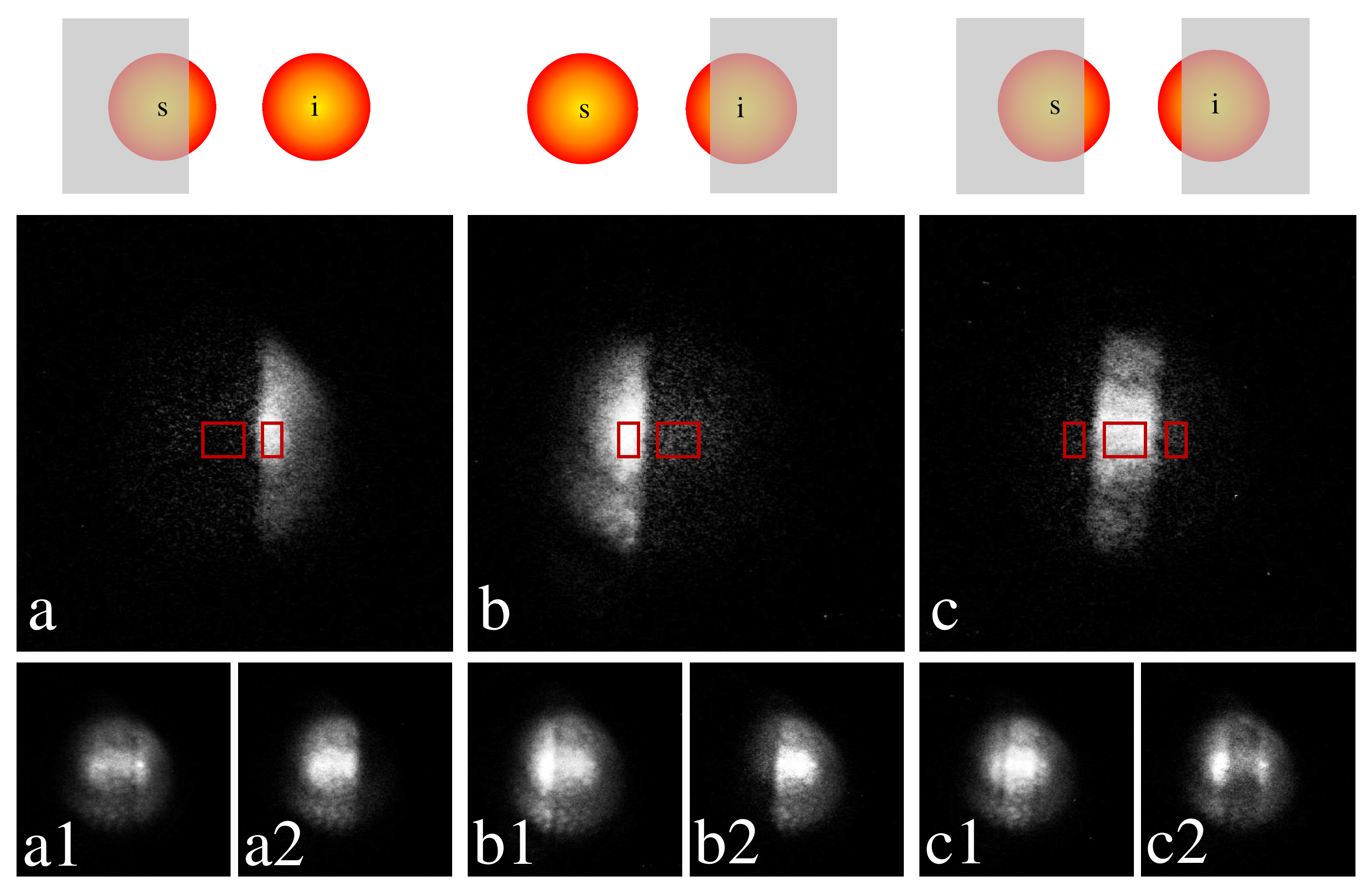}
	\caption{Experimental results of the nonlocal phase erasure of two overlapping glass plates. (a) and (b) are the heralded images of a glass plate partially inserted into the beam path of the clockwise arm of the signal photon SI and the idler photon SI, respectively. Franson interference (FI) is lost in the region covered by the glass plate (dark region). (c) is the heralded image taken when a glass plate is inserted into the clockwise arm of each SI, FI is restored only in the region where the images of the two glass plates overlap. By recording the average photons/pixel and corresponding standard deviation in the regions enclosed by the red squares and then using Eq.~(\ref{SNR}), the SNR of the images (a), (b) and (c) is calculated to be 4.05, 4.57 and 4.59 respectively. The illustration on the top shows the position of the glass plates inside the clockwise arm of the interferometer for the signal and idler beam. (a1), (b1) and (c1) are the images in the $\frac{1}{\sqrt{2}}(\ket{H}_{s}\ket{H}_{i}+\ket{V}_{s}\ket{V}_{i})$ basis (constructive interference). (a2), (b2) and (c2) are images taken in the $\frac{1}{\sqrt{2}}(\ket{H}_{s}\ket{V}_{i}+\ket{V}_{s}\ket{H}_{i})$ basis (destructive interference). The images of (a), (b) and (c) are taken by subtracting the corresponding destructive interference image from the constructive interference image. The integration time for the images is 300 seconds.}
	\label{Fig.2}
\end{figure}

Our demonstration of nonlocal phase erasure is shown in Fig.~\ref{Fig.2}. In Fig.~\ref{Fig.2}-a and b we show the images taken with coincidence imaging when a glass plate is partially inserted into the beam path of the clockwise arm of the signal photon SI (Fig.~\ref{Fig.2}-a) or the clockwise arm of the idler photon SI (Fig.~\ref{Fig.2}-b). As the thickness of the glass plate ($\sim1$~mm) is much longer than the coherence length of the SPDC photons ($\sim0.02$~mm), this makes $\Delta L_s-\Delta L_i$ also longer than the coherence length of the SPDC photons, thus eliminating FI in the region where the glass plate is present. Therefore, in the region of the beam blocked by the glass plate, we observe an equal amount of detected photons in the two basis elements $\frac{1}{\sqrt{2}}(\ket{H}_{s}\ket{H}_{i}+\ket{V}_{s}\ket{V}_{i})$ (Fig.~\ref{Fig.2}-a1 and b1) and $\frac{1}{\sqrt{2}}(\ket{H}_{s}\ket{V}_{i}+\ket{V}_{s}\ket{H}_{i})$ (Fig.~\ref{Fig.2}-a2 and b2). However, when we insert two identical glass plates, one in each SI, $\Delta L_s-\Delta L_i$ is restored to its original value in the region where the images of the two glass plates overlap, thus recovering FI and erasing the presence of the two glass plates in this region only, as seen in Fig.~\ref{Fig.2}-c. To make a quantitative comparison of the images, we make use of the signal-to-noise ratio (SNR), defined as
\begin{equation}
	\text{SNR} = \frac{1}{\sigma}|\bar{I}_{\text{in}}-\bar{I}_{\text{out}}|,
\label{SNR}
\end{equation}
where $\bar{I}_{\text{in}}$ and $\bar{I}_{\text{out}}$ are the average intensity values of the reconstructed	image, inside and outside the object profile, respectively, and $\sigma:=\sigma(\bar{I}_{\text{in}}-\bar{I}_{\text{out}})$ is the standard deviation in the intensity difference. The SNR for the image of a single glass plate in Fig.~\ref{Fig.2}-a and b are 4.05 and 4.57 respectively. The SNR for the overlap region of the two glass plates in Fig.~\ref{Fig.2}-c is 4.59, implying a full recovery of FI in that region.

\begin{figure}[htbp]
	\centering \includegraphics[width=0.48\textwidth]{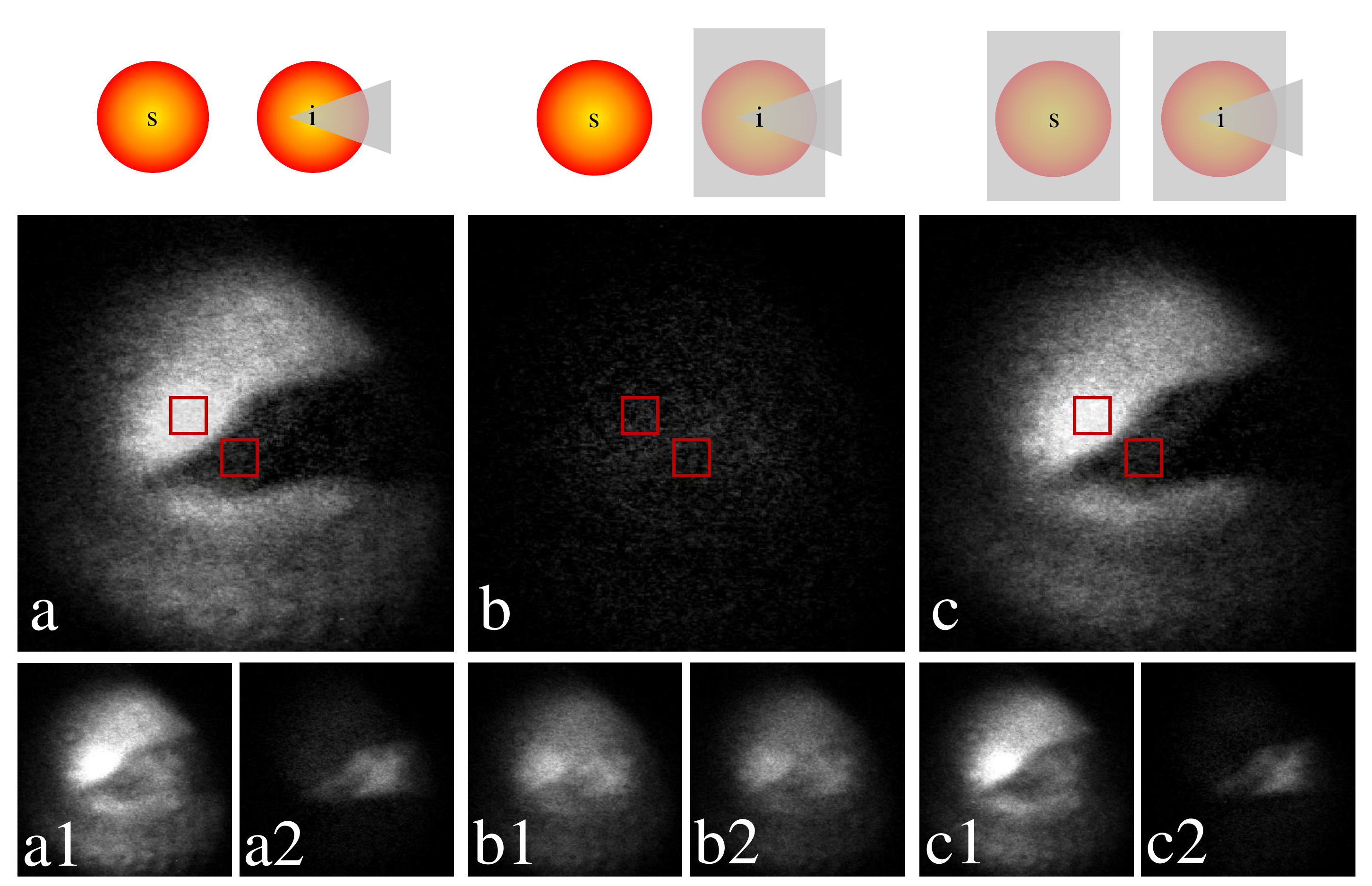}
	\caption{Experimental results demonstrating nonlocal phase correction. (a) is the heralded image of a glass shard inserted into the clockwise arm of the idler photon SI. When a glass plate is inserted behind the glass shard, FI is lost and the glass shard can no longer be seen, as shown in (b). In (c), the glass shard reappears when an identical glass plate is inserted in the clockwise arm of the signal photon SI. By recording the average photons/pixel and corresponding standard deviation in the regions enclosed by the red squares and then using Eq.~(\ref{SNR}), the SNR of the images (a), (b) and (c) is calculated to be 7.42, 0.124 and 8.57 respectively. Illustration on the top shows the position of the glass plates inside the clockwise arm of the interferometer for the signal and idler beam. The images (a1), (b1) and (c1) are the images taken in the $\frac{1}{\sqrt{2}}(\ket{H}_{s}\ket{H}_{i}+\ket{V}_{s}\ket{V}_{i})$ basis (constructive interference). (a2), (b2) and (c2) are images taken in the $\frac{1}{\sqrt{2}}(\ket{H}_{s}\ket{V}_{i}+\ket{V}_{s}\ket{H}_{i})$ basis (destructive interference). The images of (a), (b) and (c) are taken by subtracting the corresponding  destructive interference image from the constructive interference image. The integration time for the images is 300 seconds.}
	\label{Fig.3}
\end{figure}

Next, we show how nonlocal quantum erasure can potentially be used for background phase correction in Fig.~\ref{Fig.3}. A piece of glass shard ($\sim0.5$~mm) is inserted in the clockwise arm of the idler photon SI. As the thickness of the glass shard is larger than the coherence length of the SPDC photons, FI is lost in the region blocked by the shard and a shadow of the shard is seen on the camera as shown in Fig.~\ref{Fig.3}-a, the SNR for the glass shard is 7.42. A glass plate ($\sim1$~mm) is then inserted behind the glass shard, disrupting FI over the entire beam, and the interferometer is no longer phase sensitive. The image of the glass shard is thus lost due to this extra \qo{background} phase, as seen in Fig.~\ref{Fig.3}b, with a SNR of only 0.124. However, by inserting an identical glass plate in the clockwise arm of the signal photon SI, the phases of the two glass plates cancel and the glass shard is revealed again as seen in Fig.~\ref{Fig.3}-c, with a SNR of 8.57.

\setlength{\parindent}{0pt}
%

%\section{Conclusion}

In summary, we have demonstrated nonlocal quantum erasure of the presence of a phase object by employing FI. By employing this technique it is possible to perform the erasure irrespective of the distance between the erasing and erased phase objects as long as entanglement is maintained between the two photons. We also demonstrated how this method can be potentially used for remote phase corrections. We believe this technique will have many potential applications. \COR{These include quantum microscopy where one could measure the phase of specific features in a transparent or semitransparent biological sample or erasure of certain unwanted features from the sample. This can be done by placing a deformable mirror or a spatial light modulator on the image plane of the sample in the paired SI. This way the phase features of the sample can be measured/erased pixel by pixel by the deformable mirror or spatial light modulator.} One can also use this technique as a type of dual user authentication for quantum key distribution or other quantum cryptographic schemes. Here, two identical key cards with an imprinted phase pattern are issued to two parties (Alice and Bob, who need not know each other beforehand) sharing a quantum network like the one proposed in~\cite{Ekert1992}. When Alice inserts a key card into her interferometer, FI is lost and the QKD network can no longer function. To restore FI and the operation of the QKD network, Bob will have to insert an identical key card into his interferometer. \COR{Since Alice and Bob use the same QKD network, the authentication scheme should have the same security, up to the possibility of faking one of the cards (which becomes increasingly hard when increasing the complexity of the imprinted phase pattern). When the two key cards match, the ratio of photons detected in coincidence for constructive and destructive interference should be 1:0. When they do not match, or if an eavesdropper is present, then this ratio would become 1:1, a much more significant difference than the 25$\%$ quantum bit error rate for the BB84 protocol. Thus the authentication process would require much less photons to check for ensuring the security when compared, for instance, to the BB84 protocol. It might be interesting to further analyze the proposed authentication scheme within a future work and implement it within a specific cryptographic protocol.}

\begin{acknowledgments}
This work was supported by Canada Research Chairs (CRC), Canada Foundation for Innovation (CFI), Canada First Excellence Research Fund (CFREF), and Ontario’s Early Re- searcher Award. L. Gao thanks the National Nature Foundation for the support within the project No.11504337, and the Fundamental Research Funds for the Central Universities of China University of Geosciences (Beijing).
\end{acknowledgments}

\bibliographystyle{apsrev4-1}
\bibliography{references}

\end{document}